\documentclass[aps,prb,preprint,showpacs,showkeys,groupedaddress]{revtex4}
\usepackage{graphicx}
\begin{document}

\title{Transport mechanism through metal-cobaltite interfaces}

\author{C. Acha}
\thanks{corresponding author (acha@df.uba.ar)}
\address{Laboratorio de Bajas
Temperaturas - Departamento de F\'{\i}sica - FCEyN - Universidad de
Buenos Aires and IFIBA - CONICET,  Pabell\'on I, Ciudad
Universitaria, C1428EHA Buenos Aires, Argentina.}
\author{A. Schulman}
\thanks{University of Buenos Aires, Conicet and Erasmus scholarships. Present address: Correlated Electronics Group, AIST Tsukuba Central 5, 1-1-1 Higashi, Tsukuba, Ibaraki 305-8565, Japan.}
\affiliation{Laboratorio de Bajas Temperaturas - Departamento de
F\'{\i}sica - FCEyN - Universidad de Buenos Aires and IFIBA -
CONICET,  Pabell\'on I, Ciudad Universitaria, C1428EHA Buenos Aires,
Argentina.}
\address{Laboratoire des Mat\'{e}riaux et du G\'{e}nie Physique, UMR 5628 CNRS-UDG-Grenoble
INP, Minatec 3, Parvis Louis N\'{e}el, CS 50257, 38016 Grenoble,
Cedex 1, France.}
\author{M. Boudard}
\address{Laboratoire des Mat\'{e}riaux et du G\'{e}nie Physique, UMR 5628 CNRS-UDG-Grenoble
INP, Minatec 3, Parvis Louis N\'{e}el, CS 50257, 38016 Grenoble,
Cedex 1, France.}
\author{K. Daoudi}
\address{Department of Applied Physics and Astronomy, University of Sharjah,
P.O. Box 27272, Sharjah, United Arab Emirates.}
\author{T. Tsuchiya}
\address{National Institute of Advanced Industrial Science and Technology (AIST), Ibaraki 305-
8565, Japan.}

\date{\today}


\begin{abstract}
The resistive switching (RS) properties as a function of temperature
were studied for Ag/La$_{1-x}$Sr$_x$CoO$_{3}$ (LSCO) interfaces. The
LSCO is a fully-relaxed 100 nm film grown by metal organic
deposition on a LaAlO$_3$ substrate. Both low and a high resistance
states were set at room temperature and the temperature dependence
of their current-voltage (IV) characteristics was measured taking
care to avoid a significant change of the resistance state. The
obtained non-trivial IV curves of each state were well reproduced by
a circuit model which includes a Poole-Frenkel element and two ohmic
resistances. A microscopic description of the changes produced by
the RS is given, which enables to envision a picture of the
interface as an area where conductive and insulating phases are
mixed, producing Maxwell-Wagner contributions to the dielectric
properties.

\end{abstract}



\maketitle


Resistive switching (RS) has focused much attention in recent years
for being the mechanism on which one of the most promising
non-volatile memory device (RRAM or Memristor) is
based.~\cite{Yang13} Among all the investigated RRAM devices, based
on a variety of materials and different mechanisms\cite{Waser09},
those related to metal/perovskite oxides usually present a bipolar
switching type\cite{Sawa08}, whose origin can be associated with a
voltage-driven oxygen vacancy migration near the
interface.~\cite{Rozenberg10} Depending on their metal/oxide
interface, different scenarios have been considered to explain which
are the particular effects at the atomic level produced by this ion
reconfiguration. These scenarios were elaborated by analyzing the
electrical transport behavior of the switching device, in order to
reveal the mechanism behind their properties.~\cite{Sze06}
Particularly, their isothermal current-voltage (IV) characteristics
are useful to distinguish if the conduction of the device is related
to the existence of an ohmic behavior ($I \sim V$), a space charge
limited conduction (SCLC, $I \sim V^2$), or Poole-Frenkel (PF),
Fowler-Nordheim (FN) or Schottky (Sch) emissions [$I \sim
\exp(V^n)$]. For example, an electric-field-trap-controlled SCLC was
proposed to explain the RS behavior of
Ag/La$_{0.7}$Ca$_{0.3}$MnO$_{3-\delta}$ interfaces~\cite{Shang06},
while for Au/YBa$_2$Cu$_3$O$_{7-\delta}$, a PF conduction in a
variable-range hopping scenario was considered, with a
pulse-controlled-trap energy level.~\cite{Schulman15}

La$_{1-x}$Sr$_x$CoO$_{3-\delta}$ (LSCO) is one of the perovskite
oxides with high oxygen mobility were the existence of RS has been
put in evidence \cite{Hamaguchi06,Fu14}. In particular Fu et
al.\cite{Fu14} have recently investigated the RS on
Ag/La$_{0.5}$Sr$_{0.5}$CoO$_{3-\delta}$ 200nm thick films grown by
pulsed laser deposition on Pt/Ti/SiO2/Si substrates, showing that
their transport mechanism was controlled by a SCLC with traps
exponentially distributed in energy ($I \sim V^m$, with
m=1;2;$\geq$7, depending on the different SCLC regimes explored).

In this work, RS and isothermal IV characteristics of Ag/LSCO
interfaces are reported in two well differentiated low and high
resistance states (LRS and HRS) in order to point out their
microscopic differences. We propose an equivalent circuit model with
a PF element and two ohmic resistances that reproduces the
non-trivial electrical behavior of this memristor. Besides, we found
that the metal/oxide interface may be composed by a nanometric
mixture of conducting and insulating zones, yielding to an
enhancement of the contact dielectric constant due to Maxwell-Wagner
(MW) type contributions.

\begin{figure}
\vspace{-0mm}
\centerline{\includegraphics[angle=0,scale=0.7]{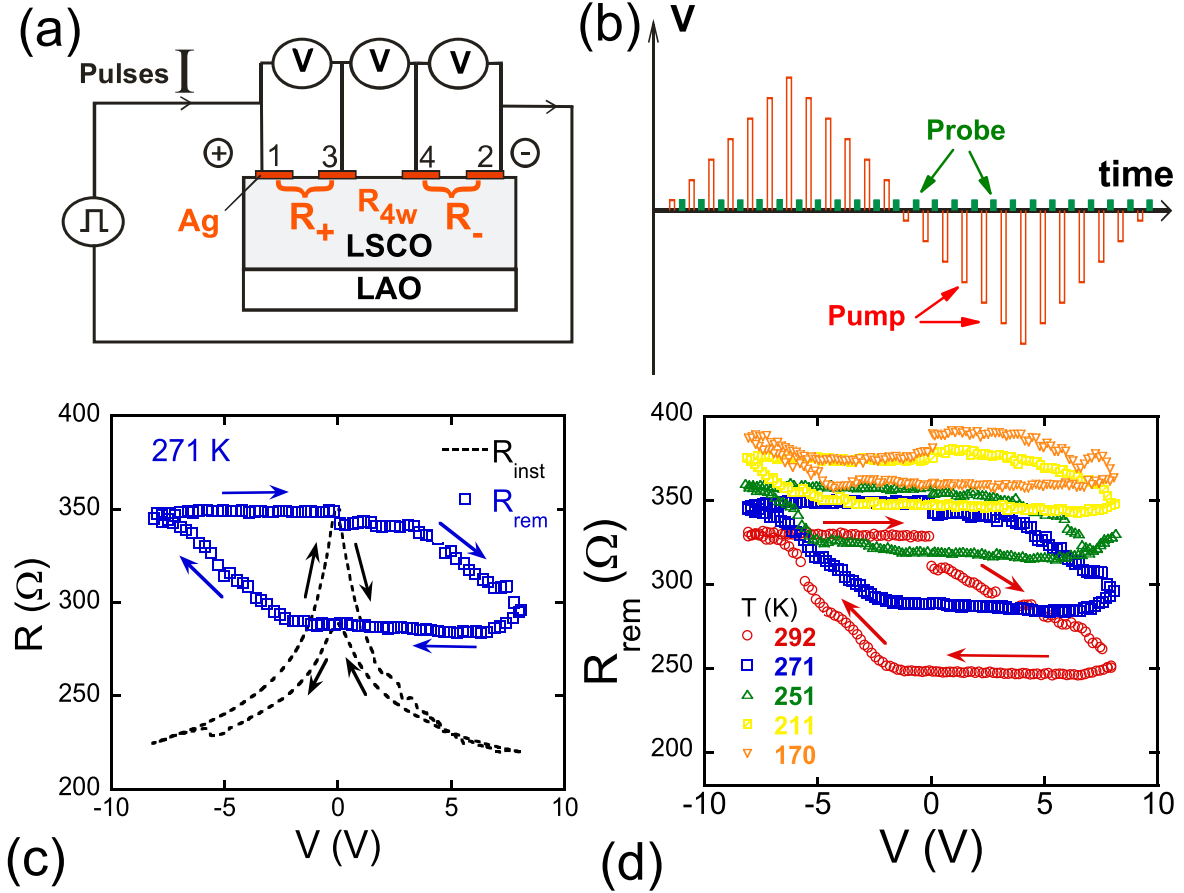}}
\vspace{-5mm}\caption{(Color online) (a) Schematic representation of
the device and its electrode configuration. (b) Voltage pump (red)
and probe (green) protocol followed to measure the IV
characteristics, the instantaneous ($R_{inst}$) and the remanent
resistance ($R_{rem}$). (c) RHSL of both $R_{inst}$ and $R_{rem}$
measured at 271 K for a $\pm$ 8 V cycle. The existence of a
non-volatile memory, two well differentiated LRS and HRS as well as
non-linear effects can be observed. The arrows indicate the
clockwise circulation of the cycle.  (d) RHSL of $R_{rem}$ at
different temperatures. } \vspace{-0mm} \label{fig:contactos}
\end{figure}

An epitaxial 100 nm thick film of La$_{0.7}$Sr$_{0.3}$CoO$_3$ (LSCO)
was grown on a 10x10x0.5 mm$^3$ LaAlO$_3$ (LAO) single crystal (001)
substrate (Crystec) by using a conventional metal organic deposition
process (with a final annealing at 1000 $^\circ$C in air). A
detailed description of the growth process, phase purity,
structural, morphological, magnetic and electric characterizations
can be found in previous works.~\cite{Daoudi10,Othmen14} To perform
electrical transport measurements, four hand-painted Ag electrodes
were done on top of the surface of the film. The electrodes,
depicted in Fig.~\ref{fig:contactos}(a), have approximately a width
of 0.75 mm and a mean separation between them of 0.6 $\pm$ 0.1 mm .

The IV characteristics of the $R_+$ interface (including the
interfacial zone plus a fraction of the bulk material between
electrodes 1-3 (see Fig.~\ref{fig:contactos}(b)) were measured at
different fixed temperatures using two synchronized Keithley 2400
precision source/measure units (SMU). While one of the SMU provides
10 ms square voltage pump pulses to electrodes 1-2, with and
amplitude following the triangular shape shown in
Fig.~\ref{fig:contactos}(b) in a $\pm$ 10 V range, and measures
current ($I$) during the last 50\% of the pulse´s period, where the
signal remains practically constant (stationary regime), the other
SMU measures de voltage drop between electrodes 1 and 3 (V13,
hereafter noted as $V$). In this way, the instantaneous resistance
of this interface can be defined as $R_{inst}=V/I$. After 300 ms of
each pump pulse, a 10 ms probe pulse corresponding to a small bias
voltage pulse (50 mV) was applied in order to measure in the same
way the remnant resistance ($R_{rem}$) of the junction (see
Fig.~\ref{fig:contactos}(b)). Then, after an additional waiting time
of 300 ms, the cycle starts again with the following pump
pulse.\cite{R-} The four wires resistance [$R_{4W}(T)$] of the LSCO
sample was measured applying a small bias current (10 $\mu$A)
(Keithley 220 current source) to electrodes 1 and 2 and measuring
V34 (Agilent 34430A nano-voltmeter). Temperature was stabilized with
a stability better that 0.3\% for each IV characterization in the 80
$<$ T $<$ 300 K range and measured with a Pt thermometer well
thermally anchored to the sample.


The existence of RS properties for our Ag/LSCO device was first
checked by applying the pulsing protocol presented in
Fig.~\ref{fig:contactos}(b) and by measuring its instantaneous and
remnant response. These results can be observed in
Fig.~\ref{fig:contactos}(c), where a typical resistance hysteresis
switching loop (RHSL) is shown at $T \simeq$ 271 K, where both, the
instantaneous non-linear resistance ($R_{inst}$) and the remnant
resistance ($R_{rem}$), are plotted as a function of the local
voltage $V$ applied  during the pulsing protocol. A cyclic evolution
of $R_{rem}$ can be observed, between low and high values for pulses
in the $\pm$ 8 V range and with a clockwise circulation. Within the
voltage amplitude of these pulses, a reproducible ratio $\alpha =
\frac{R_{rem}^{High}}{R_{rem}^{Low}} \simeq$ 1.3 was obtained. As
shown in Fig.~\ref{fig:contactos}(d9, a similar behavior is obtained
for the whole $T$ range studied (80 K $<$ T $<$ 300 K), with a
reduction of $\alpha$ to $\simeq$ 1.08 with decreasing $T$.

\begin{figure}
\vspace{-0mm}
\centerline{\includegraphics[angle=0,scale=0.7]{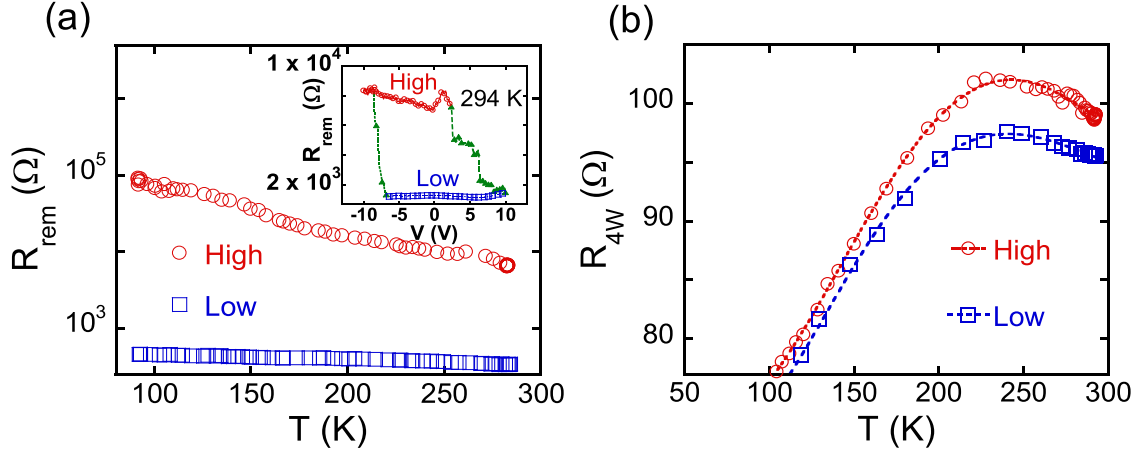}}
\vspace{-5mm}\caption{(Color online) (a) Remnant resistance
$R_{rem}$ measured during each low-voltage IV cycle at different
temperatures for each room-temperature-programmed LRS and HRS. The
inset shows a typical RHSL of $R_{rem}$ performed at room
temperature to switch from one state to the other. (b) $R_{4W}$ vs
$T$ measured for each programmed state. } \vspace{-0mm}
\label{fig:RremvsT}
\end{figure}

In order to reveal the main conduction mechanism through the device
and particularly the microscopic differences between the LRS and the
HRS, the IV characteristics of the Ag/LSCO junction as a function of
$T$ should be measured in two stable and well differentiated states.
To program these states and to facilitate this task, a higher
$\alpha$ $\sim$ 10 was produced at room temperature ($R_{rem}^{Low}
\sim$ 1 k$\Omega$; $R_{rem}^{High} \sim$ 10 k$\Omega$) by increasing
the amplitude of the negative pulses up to -10 V (see the inset of
Fig.~\ref{fig:RremvsT}(a).

After generating each state, the voltage amplitude of the pulses
during the IV measurement cycle was maintained sufficiently low in
order to prevent a RS that would change the microscopic
characteristics that determine the properties of the device. From
the RHSL of $R_{rem}$ shown in Fig.~\ref{fig:contactos}(c) a safe
voltage amplitude of -1.5 V $< V <$ 1.0 V was then chosen to perform
the IV characteristics at different stabilized temperatures in the
80 K $<$ T $<$ 300 K range.

As can be observed in Fig.~\ref{fig:RremvsT}(a), $R_{rem}$  was
measured for both selected states before and after each IV
measurement, for experiments performed at different temperatures. It
is clear that no significant RS was produced along the entire series
of IV measurements. The $T$ dependence of $R_{4W}$ for both states
(Fig.~\ref{fig:RremvsT}(b) shows the typical behavior measured
before the pulsing treatments.~\cite{Othmen14} Surprisingly, the
measured $R_{4W}$ is different for each contact state, although
$R_{4W}$ is measured with contacts 3 and 4 (see
Fig.~\ref{fig:contactos}(a) located at a macroscopic distance
($\sim$ 1 mm) from the pulsed contacts 1-2. The increase of
resistivity when switching from LRS to HRS is in accordance with a
reduction of the local oxygen content.~\cite{Liu16} This indicates
that the pulsing affects zones far away from the interfaces, as was
previously observed~\cite{Acha09a} in ceramic YBa$_2$Cu$_3$O$_7$
(YBCO), probably associated to the high mobility of oxygen in these
oxides~\cite{Chroneos10,Ji13} and to its relation to the mechanism
of the resistance change induced by the pulsing
treatment.~\cite{Rozenberg10}

\begin{figure} [b]
\vspace{-0mm}
\centerline{\includegraphics[angle=0,scale=0.7]{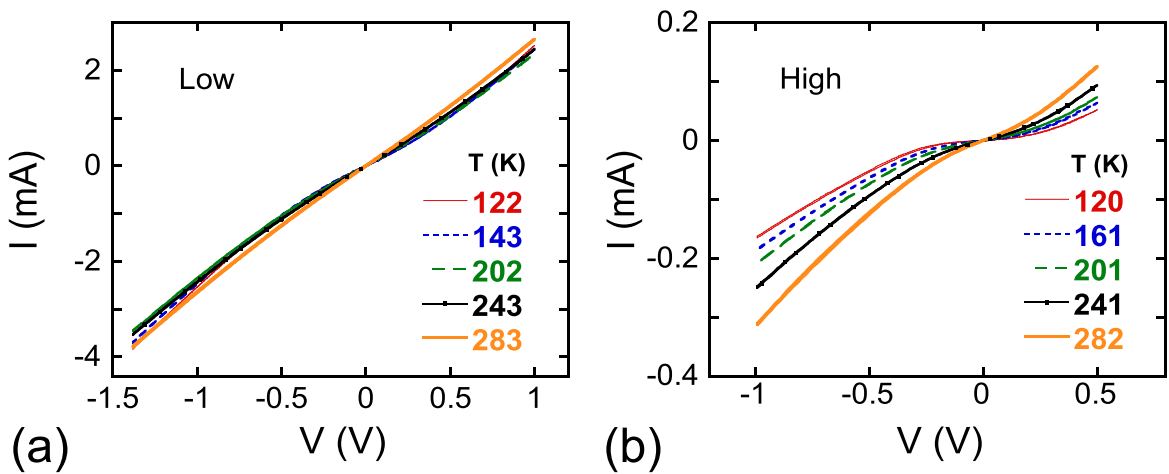}}
\vspace{-5mm}\caption{(Color online) IV characteristics at different
temperatures for (a) the LRS and (b) the HRS. Non-linear effects are
clearly visible as well as a non-rectifying behavior.} \vspace{-0mm}
\label{fig:IV_Txx_compara}
\end{figure}

\begin{figure} [t]
\vspace{-0mm}
\centerline{\includegraphics[angle=0,scale=0.65]{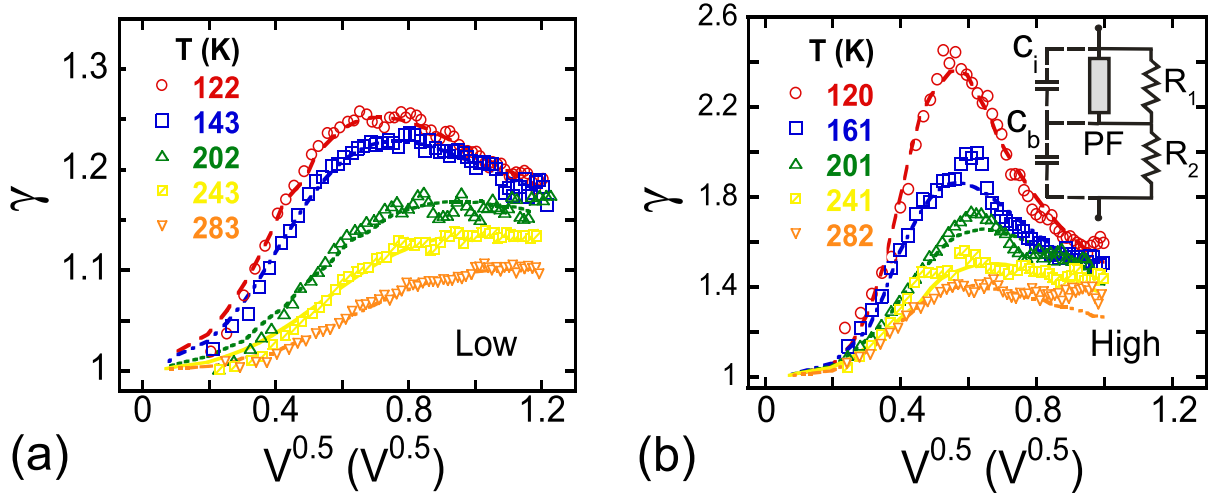}}
\vspace{-5mm}\caption{(Color online) $\gamma$ as a function of
$V^{1/2}$ at different temperatures for the programmed (a) LRS and
(b) HRS. The dashed-lines correspond to fits of the experimental
data using Eq.~\ref{eq:modelo1}. The inset is a schematic of the
circuit elements used to model the IV characteristics.}
\vspace{-0mm} \label{fig:gamma}
\end{figure}

The IV characteristics for both selected states are shown in
Fig.~\ref{fig:IV_Txx_compara}. Both states show non-linear effects.
The LRS shows a small $T$ dependence, while the HRS shows a higher
sensitivity to it. A Schottky mechanism can be ruled out as no
rectifying behavior can be observed. Details of the IV
characteristics are more readily apparent if one considers the
derivative $\gamma = dLn(I) / dLn(V)$ plotted as a function of
$V^{1/2}$ (see Fig.~\ref{fig:gamma}).~\cite{Bozhko02} For clarity,
only the increasing negative voltage part of the IV measurement
cycle was used to plot these curves (taking the absolute value of
V). Indeed, the typical conduction mechanisms through metal-oxide
interfaces can be easily determined by analyzing these $\gamma$ vs
$V^{1/2}$ curves. This is the case for an ohmic, a Langmuir-Child or
a SCLC conduction, that will all show a constant $\gamma$ ($=$ 1,
1.5 or 2, respectively), or for a Schottky (Sch) or a Poole-Frenkel
(PF) behavior that will be represented by a straight line differing
in the intercept (0 for Sch, 1 for PF).~\cite{Sze06} Our results are
described by a more complex $\gamma$ vs. $V^{1/2}$ curve as it can
be observed in Fig.~\ref{fig:gamma}. Both states show for low
voltages a $\gamma$ tending to a constant value close to 1 (more
clearly defined for the HRS), then a small linear region limited by
the appearance of a maximum (more prominent at low temperatures and
for the HRS) and a final gradual decrease.

This complex behavior can be reproduced by considering the circuit
elements depicted in the inset of Fig.~\ref{fig:gamma}(b). This
representation of the memristor by a similar arrangement of circuit
elements was already done in previous papers to describe the
dynamical behavior of metal-YBCO interfaces~\cite{Acha11}, and the
IV characteristics of metal-manganite
junctions.~\cite{Marlasca13,capacitors}

We would like to note here that the data reported by Fu et
al.~\cite{Fu14} on the IV characteristics of
Ag/La$_{0.5}$Sr$_{0.5}$CoO$_{3-\delta}$ interfaces presents a
similar $\gamma(V)$ behavior to the one reported here. Their
$\gamma$ evolves gradually with increasing V from 1 to a maximum
value slightly higher than 2 and then decreases smoothly with
further increasing V. Here, by considering the almost linear parts
of the $\gamma(V^{0.5})$ curve and by discarding a Schottky
conduction (as mentioned, due to the lack of rectifying behavior),
the non-linear element was associated with a PF conduction. A
parallel ohmic element ($R_1$) is needed to reproduce the
quasi-ohmic dependence observed at low voltages, while the ohmic
element in series ($R_2$) should be considered to limit the
non-linear conduction, producing the maximum and the concomitant
decrease of $\gamma$ observed for higher voltages. In this way, the
following equations represents the IV characteristics of the
proposed representation for the Ag/LSCO
memristor:~\cite{Simmons67,Sze06}

\vspace{-4mm}

\begin{equation}
\label{eq:modelo1} I = A ~(V-IR_2) \exp[\frac{B(V-IR_2)^{1/2}}{k_B
T}] + \frac{V-IR_2}{R_1},
\end{equation}

\noindent with
\begin{equation}
\label{eq:modelo2} A = \frac{\exp(-\frac{\phi_B}{k_B T})}{R_{ox}}, B
= \frac{q^{3/2}}{(\pi \epsilon^{'} d)^{1/2}}, R_{ox} =
\frac{d}{Sqn_0\mu},
\end{equation}

\noindent where $T$ is the temperature, $k_B$ the Boltzmann
constant, $\phi_B$ the trap energy level, $q$ the electron's charge,
$S$ the conducting area, $\epsilon^{'}$,$n_0$ and $ \mu$ the real
part of the dielectric constant, the density of carriers and their
mobility in the oxide, respectively, and $d$ the distance where the
voltage drop is produced.

The dashed-lines plotted in Fig.~\ref{fig:gamma} were obtained by
fitting numerically the IV curves, by using the implicit
Eq.~\ref{eq:modelo1}. The non-trivial experimental $\gamma(V^{0.5})$
dependence is very well reproduced by the proposed model. Only a
slight difference can be noted for the highest voltages in the HRS,
where $\gamma$ seems to saturate at a value close to 1.5 instead of
the expected 1. This is possibly indicating the existence of a more
complex process in parallel with $R_2$, like a $T$-independent
Langmuir-Child conduction due to ballistic carriers through vacuum
voids or oxygen-depleted zones~\cite{Sze06,Srisonphan12}. More work
is needed to address these deviations, which are beyond the scope of
the present paper.

\begin{figure}
\vspace{-0mm}
\centerline{\includegraphics[angle=0,scale=0.65]{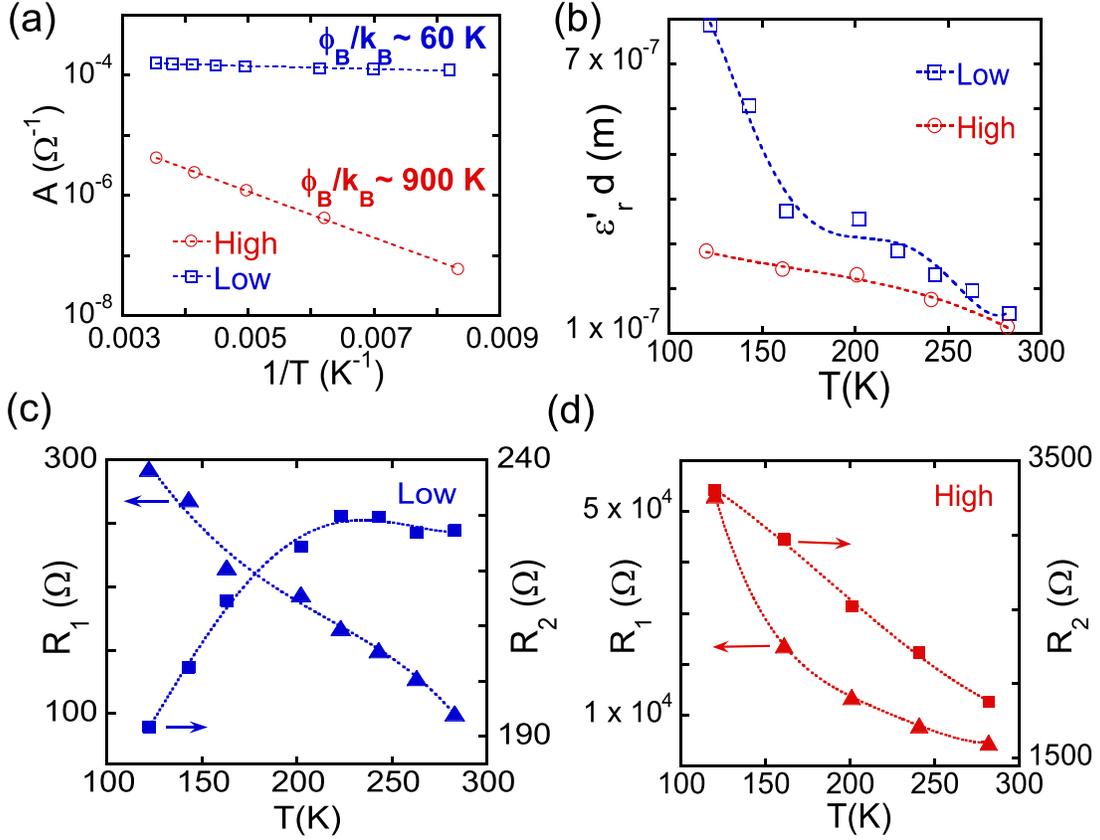}}
\vspace{-5mm}\caption{(Color online) Parameters obtained from the
fits of the experimental data presented in
Fig.~\ref{fig:IV_Txx_compara} for the LRS and the HRS by
Eq.~\ref{eq:modelo1}. (a) Semilogaritmic plot of A vs 1/T.
Dashed-lines are fits of the data by using Eq.~\ref{eq:modelo2}. (b)
$\epsilon_r^{'}~d$ vs T derived from the B fitting-parameter by
using Eq.~\ref{eq:modelo2}. The dashed lines are guides to the eye.
(c) and (d) $R_1(T)$ and $R_2(T)$ for the LRS and the HRS,
respectively. Dashed-lines are guides to the eye.} \vspace{-0mm}
\label{fig:param}
\end{figure}

We reduced the four fitting parameters $A$, $B$, $R_1$ and $R_2$ to
three by considering that $R_1$ and $R_2$ can be related to the
measured $R_{rem}(T)$ using the fact that, for sufficiently low
voltages, $R_{rem} \simeq R_1/(AR_1+1) + R_2$ (see
Ref.~\onlinecite{Supple1} for a list of typical parameters obtained
for each state). Fig.~\ref{fig:param} shows the $A$, $B^{-2}$,
$R_1$, and $R_2$ parameters obtained from the fits. For both
resistance states, a straight line is obtained when plotting Ln(A)
vs 1/T, indicating that the exponential relation of $A$ with 1/T
derived from the PF conduction is valid (see Eq.~\ref{eq:modelo2}).
The slope, associated with the trap potential $\phi_B$, indicates
the presence of traps [$\phi_B/(k_B) \sim$ 900 K] in the interfacial
zone of the oxide for the HRS, while the flatness of the curve for
the LRS points out practically their absence [$\phi_B/(k_B) \sim$ 60
K]. Similar $\phi_B/(k_B)$ values ($\sim$ 700-1200 K) were obtained
for Au/YBCO interfaces after pulsing treatments.~\cite{Schulman15}
In the same way, by analyzing the intercept, it can be derived that
$R_{ox}^{HRS} \simeq 2R_{ox}^{LRS}$. This indicates that not only
deeper traps are created when switching from LRS to HRS but also
that the resistivity of the interfacial zone, where the PF
conduction develops, increases, which is probably associated with an
overall reduction of the local oxygen content.

The $B$ parameter is plotted in Fig.~\ref{fig:param}(b) as $B^{-2}$
times $q^3/(\pi \epsilon_0)$ vs T, in order to represent, according
to Eq.~\ref{eq:modelo2}, the $T$ dependence of $\epsilon_r^{'}~d$,
where $\epsilon_r^{'} = \epsilon^{'} / \epsilon_0 $ (with
$\epsilon_0$ being the vacuum permittivity). As the distance $d$,
associated with the interfacial zone, is expected to be $T$
independent, and to remain practically unchanged for both states, as
was reported for Au/YBCO~\cite{Schulman15}, the $T$ dependence of
$\epsilon_r^{'}$ can be directly inferred from
Fig.~\ref{fig:param}(b). It can be observed that $\epsilon_r^{'}$ in
the LRS increases with decreasing $T$, following a step-like shape,
while in the HRS it shows a tendency to saturate for temperatures
below 200 K. This non-usual dependence of $\epsilon_r^{'}$ for a
paraelectric oxide~\cite{Silverman63} was previously observed in the
case of mixtures of conducting and insulating regions, which are in
many cases generated intrinsically in metal-oxide
interfaces~\cite{Lunkenheimer02}. This mixture of phases may give
rise to a Maxwell-Wagner (MW)
effect~\cite{Lunkenheimer02,Lunkenheimer10}, where the charge
carriers accumulate at the border of the conducting regions, which
act as very small parallel-plate capacitors, representing a very
high capacitance. In this particular case, by describing the sample
as composed by an interfacial region in series with the bulk (each
material with different conductivity and dielectric constant), it
was shown that the dielectric constant has a frequency dependence
described by the Debye relaxation laws, with a static value
($\epsilon_s^{'}$) higher than the one at high frequencies
($\epsilon_\infty^{'}$) and a characteristic time ($\tau$) that
follows the $T$ dependence of the resistivity of the bulk. A step in
$\epsilon_r^{'}$ to a higher value (a shift from
$\epsilon_\infty^{'}$ to $\epsilon_s^{'}$) is then predicted as
$\tau$ is reduced when decreasing $T$. This is particularly what it
can be observed in Fig.~\ref{fig:param}(b), as in the LRS
$\epsilon_r^{'}$ increases in correspondence to the decrease of
$R_2$ (see Fig.~\ref{fig:param}(c), which represents the electrical
connection between the interfacial zone and the bulk. In contrast,
as $R_2$ remains insulating in the HRS (see Fig.~\ref{fig:param}(d),
the obtained $\epsilon_r^{'}$ is weakly $T$ dependent.

Finally, $R_1$ becomes 100 times more insulating in the HRS while
$R_2$, mimics the $T$ dependence of the $R_{4W}$ resistance in the
LRS, but becomes semiconducting-like with a 10 times increase in the
HRS. This similarity between $R_2(T)$ and $R_{4W}$(T), only
differing in a geometric constant that amplifies its value, was
previously seen in metal interfaces with YBCO\cite{Acha09a} and
manganites\cite{Quintero07}, pointing out the generality of this
process.

All these results lead us to consider that, independently of the
initial oxygen content of the oxide, the Ag/LSCO interfacial region
has become an oxygen-depleted zone, where the electrical transport
properties are dominated by the presence of charge carrier's traps
(PF) in parallel with insulating regions ($R_1$). These interfacial
characteristics may represent a common feature of other
metal/perovskite oxide interfaces, like Au/YBCO, were a similar
behavior was observed\cite{Schulman15}, probably associated with the
capacity of the metallic electrode to generate in the oxide the
coexistence of conducting and insulating regions, yielding probably
to a granular structure. This scenario would be similar to the one
existing intrinsically for phase-separated manganites, where the
coexistence of nanometric conducting and insulating regions is a
feature proper of their phase separation
characteristics.~\cite{Dagotto01} In fact, their dielectric
properties showed a behavior compatible with a MW description, as
was already reported.~\cite{Garba06,Rivas06,Lunkenheimer10}


In summary, the $T$ dependence of the RS properties of a Ag/LSCO
interface was reported. A detailed study of the IV characteristics
at different temperatures for both LRS and HRS is presented,
allowing the development of an equivalent circuit model based on a
non-linear PF element and two ohmic resistances. This model captures
the non-trivial electrical behavior of the Ag/LSCO junction that we
believe gives a general representation of other metal/perovskite
oxide memristors. This model also allows to envision details of the
microscopic constitution of the interfacial zone and to understand
which are the specific changes produced upon the pulsing treatments.

We acknowledge financial support from CONICET PIP12-14(0930),
PICT13-0788 and UBACyT14-17(20020130100036BA).

\vspace{-0mm}

\bibliography{bibRRAM}

\begin{thebibliography}{10}

\bibitem{Yang13}
J.~J. Yang, D.~B. Strukov, and D.~R. Stewart.
\newblock {\em Nature Nanotechnology}, 8:13, 2013.

\bibitem{Waser09}
R.~Waser, R.~Dittmann, G.~Staikov, and K.~Szot.
\newblock {\em J. Appl. Phys.}, 78:6113, 2009.

\bibitem{Sawa08}
A.~Sawa.
\newblock {\em Materials Today}, 11:28, 2008.

\bibitem{Rozenberg10}
M.~J. Rozenberg, M.~J. S\'anchez, R.~Weht, C.~Acha, F.~Gomez-Marlasca, and
  P.~Levy.
\newblock {\em Phys. Rev. B}, 81:115101, 2010.

\bibitem{Sze06}
S.~M. Sze and K.~K. Ng.
\newblock {\em Physics of Semiconductor Devices}.
\newblock John Wiley \& Sons, 2006.

\bibitem{Shang06}
D.~S. Shang, Q.~Wang, L.~D. Chen, R.~Dong, X.~M. Li, and W.~Q. Zhang.
\newblock {\em Phys. Rev. B}, 73:245427, 2006.

\bibitem{Schulman15}
A.~Schulman, L.~F. Lanosa, and C.~Acha.
\newblock {\em Journal of Applied Physics}, 118:044511, 2015.

\bibitem{Hamaguchi06}
M.~Hamaguchi, K.~Aoyama, S.~Asanuma, Y.~Uesu, and T.~Katsufuji.
\newblock {\em Appl. Phys. Lett.}, 88:142508, 2006.

\bibitem{Fu14}
Y.~J. Fu, F.~J. Xia, Y.~L. Jia, C.~J. Jia, J.~Y. Li, X.~H. Dai, G.~S. Fu, B.~Y.
  Zhu, and B.~T. Liu.
\newblock {\em Applied Physics Letters}, 104(22):223505, 2014.

\bibitem{Daoudi10}
K.~Daoudi, T.~Tsuchiya, T.~Nakajima, A.~Fouzri, and M.~Oueslati.
\newblock {\em J. Alloys Compd.}, 506:483, 2010.

\bibitem{Othmen14}
Z.~Othmen, A.~Schulman, K.~Daoudi, M.~Boudard, C.~Acha, H.~Roussel,
  M.~Oueslati, and T.~Tsuchiya.
\newblock {\em Applied Surface Science}, 306:60, 2014.

\bibitem{R-}
The same studies can be performed for the $R_-$ interface just by following the
  same steps but measuring the voltage drop V42 instead of V13. As both
  electrodes showed qualitatively similar results, only those for $R_+$ are
  presented here.

\bibitem{Liu16}
Bin Liu, Guiju Liu, Honglei Feng, Chao Wang, Huaiwen Yang, and Yiqian Wang.
\newblock {\em Materials and Design}, 89:715, 2016.

\bibitem{Acha09a}
C.~Acha and M.~J. Rozenberg.
\newblock {\em J. Phys.: Condens. Matter}, 21:045702, 2009.

\bibitem{Chroneos10}
A.~Chroneos, R.V. Vovk, I.L. Goulatis, and L.I. Goulatis.
\newblock {\em Journal of Alloys and Compounds}, 494(1–2):190 -- 195, 2010.

\bibitem{Ji13}
Ho-Il Ji, Jaeyeon Hwang, Kyung~Joong Yoon, Ji-Won Son, Byung-Kook Kim, Hae-Won
  Lee, and Jong-Ho Lee.
\newblock {\em Energy Environ. Sci.}, 6:116--120, 2013.

\bibitem{Bozhko02}
A.~Bozhko, M.~Shupegin, and T.~Takagi.
\newblock {\em Diamond and Related Materials}, 11:1753, 2002.

\bibitem{Acha11}
C.~Acha.
\newblock {\em J.Phys.D: Appl.Phys.}, 44:345301, 2011.

\bibitem{Marlasca13}
F.~Gomez-Marlasca, N.~Ghenzi, A.~G. Leyva, C.~Albornoz, D.~Rubi, P.~Stoliar,
  and P.~Levy.
\newblock {\em Journal of Applied Physics}, 113(14):144510, 2013.

\bibitem{capacitors}
Although the measurements were performed at low frequencies, in a stationary
  regime, the capacitances of the interfacial zone ($C_i$) and the bulk ($C_b$)
  were also considered in order to take into account the dielectric behavior of
  the device.~\cite{Lunkenheimer02}.

\bibitem{Simmons67}
J.~G. Simmons.
\newblock {\em Phys. Rev.}, 155:657, 1967.

\bibitem{Srisonphan12}
Siwapon Srisonphan, Yun~Suk Jung, and Hong~Koo Kim.
\newblock {\em Nat Nano}, 7, 2012.

\bibitem{Supple1}
See supplemental material at [URL will be inserted by AIP] for a list of the
  typical parameters obtained for both states.

\bibitem{Silverman63}
B.~D. Silverman and R.~I. Joseph.
\newblock {\em Phys. Rev.}, 129:2062, 1963.

\bibitem{Lunkenheimer02}
P.~Lunkenheimer, V.~Bobnar, A.~V. Pronin, A.~I. Ritus, A.~A. Volkov, and
  A.~Loidl.
\newblock {\em Phys. Rev. B}, 66:052105, 2002.

\bibitem{Lunkenheimer10}
P.~Lunkenheimer, S.~Krohns, S.~Riegg, S.G. Ebbinghaus, A.~Reller, and A.~Loidl.
\newblock {\em The European Physical Journal Special Topics}, 180:61, 2010.

\bibitem{Quintero07}
M.~Quintero, P.~Levy, A.~G. Leyva, and M.~J. Rozenberg.
\newblock {\em Phys. Rev. Lett.}, 98:116601, 2007.

\bibitem{Dagotto01}
E.~Dagotto, T.~Hotta, and A.~Moreo.
\newblock {\em Physics Reports}, 344:1, 2001.

\bibitem{Garba06}
G.~Garbarino, C.~Acha, P.~Levy, T.~Y. Koo, and S-W. Cheong.
\newblock {\em Phys. Rev. B}, 74:100401(R), 2006.

\bibitem{Rivas06}
J.~Rivas, J.~Mira, B.~Rivas-Murias, A.~Fondado, J.~Dec, W.~Kleemann, and M.~A.
  Se{\~n}aris-Rodr{\'i}guez.
\newblock {\em Applied Physics Letters}, 88(24):242906, 2006.

\end{thebibliography}

\end{document}